\documentclass[letterpaper, 12pt]{article}

\usepackage{times}
\usepackage{fullpage}
\usepackage{amsmath}
\usepackage{setspace}

\usepackage{graphicx}
\usepackage{pstricks}

\usepackage{subfigure}

\usepackage{array}

\graphicspath{ {picture/} }

\title{Hamamatsu R1584 PMT Modifications}
\author{Wenliang Li\footnote{li479@uregina.ca}, Garth Huber\footnote{huberg@uregina.ca}, Keith Wolbaum \\~ \\ University of Regina, Regina, SK, S4S-0A2 Canada}

\begin{document}
\maketitle

\setlength\arraycolsep{0.1em}

\doublespacing

\begin{abstract}
Four Hamamatsu H6528 Photomultiplier Tube (PMT) assemblies were purchased by the University of Regina, to be used on the SHMS Heavy Gas \v{C}erenkov detector. Despite the excellent gain, the H6528 signal output has two disturbing characteristics: discharges and ringing tails. In this report, we offer solutions to overcome these issues.
\end{abstract}

\section{Introduction}

Four 5$^{\prime\prime}$ Hamamatsu\footnote{HAMAMATSU Photonics K.K. [IR] 325-6, Sunayama-cho, Naka-ku, Hamamatsu City, Shizuoka Pref., 430-8587, Japan. Website: www.hamamatsu.com} H6528 Photomultiplier Tube (PMT) assemblies were purchased by the University of Regina in May, 2011. These detectors are used on the Heavy Gas \v{C}herenkov detector, which is part of the ongoing Hall C 12~GeV upgrade program at Jefferson Lab.

\begin{figure}[t]
\centering
\includegraphics[width=0.7\textwidth]{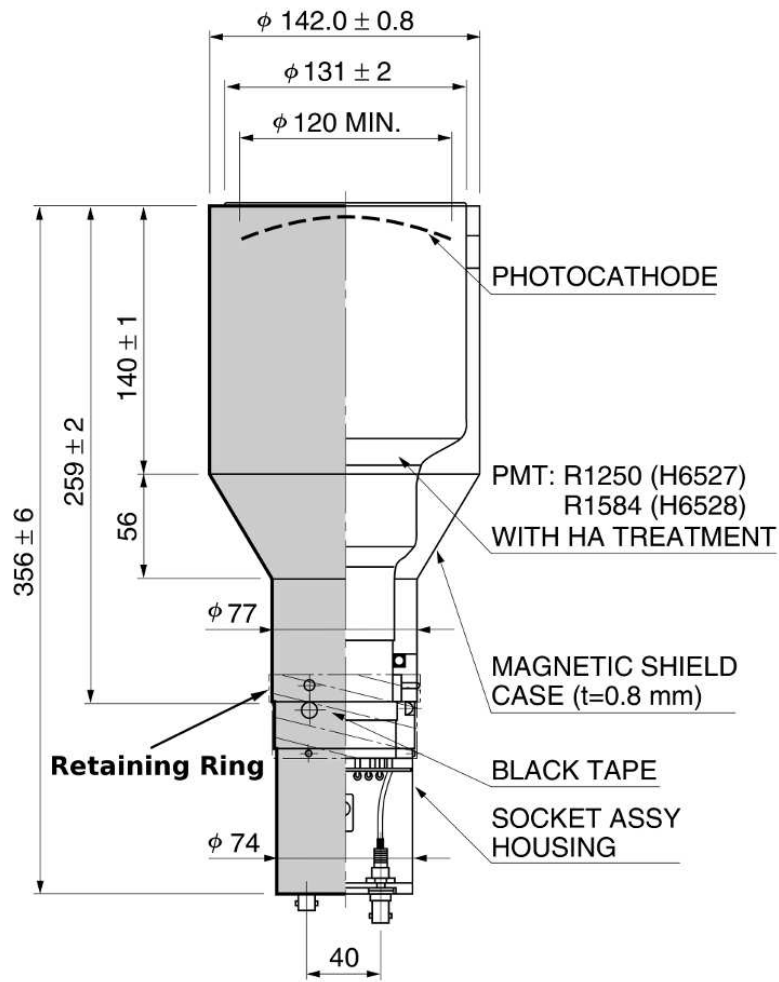} 
\caption{Hamamatsu H6528 PMT assembly schematic diagram~\cite{hamamatsu_cataloge}.}
\label{pic:R1584}
\end{figure}

The H6528 PMT assembly consists of three parts: a R1584 PMT, a magnetic shield ($\mu$-metal shield) and an electronic base, shown in Fig~\ref{pic:R1584}. Note that the $\mu$-metal shield is coated with electrical insulation paint; the housing of the electronic base is made of aluminum. The maximum operating and recommended voltages are 3000~V and 2000~V according to the specification. The detector gains were measured at the University of Regina, the results were higher than those from the factory test sheet~\cite{alex}.

\begin{figure}[t]
\centering
\subfigure[Discharge Signal \label{pic:discharge}]{\includegraphics[width=0.46\textwidth]{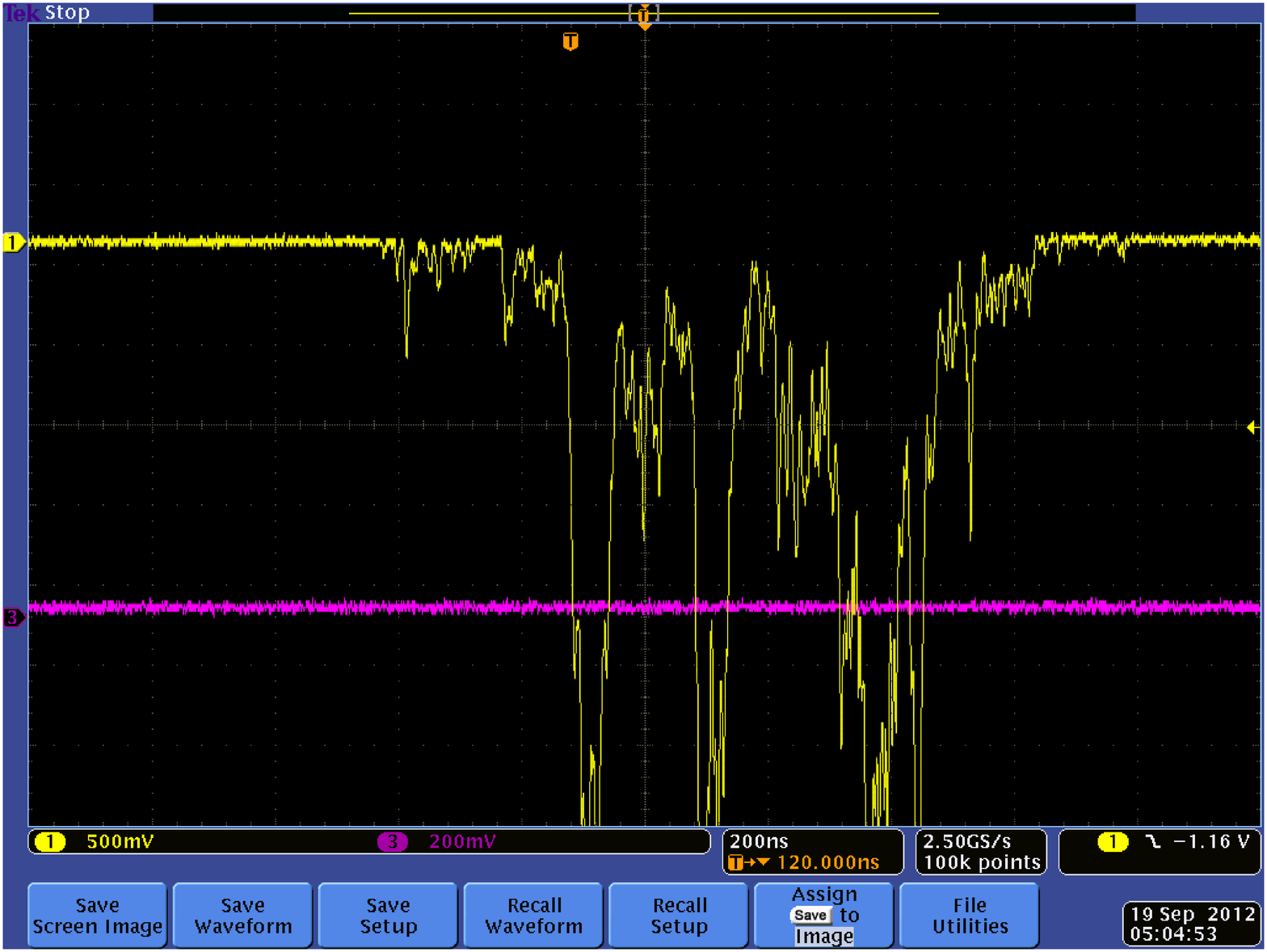}} \quad
\subfigure[Ringing Tails \label{pic:ringing}]{\includegraphics[scale=0.385, trim = 0mm 53mm 5.5mm 53mm, clip]{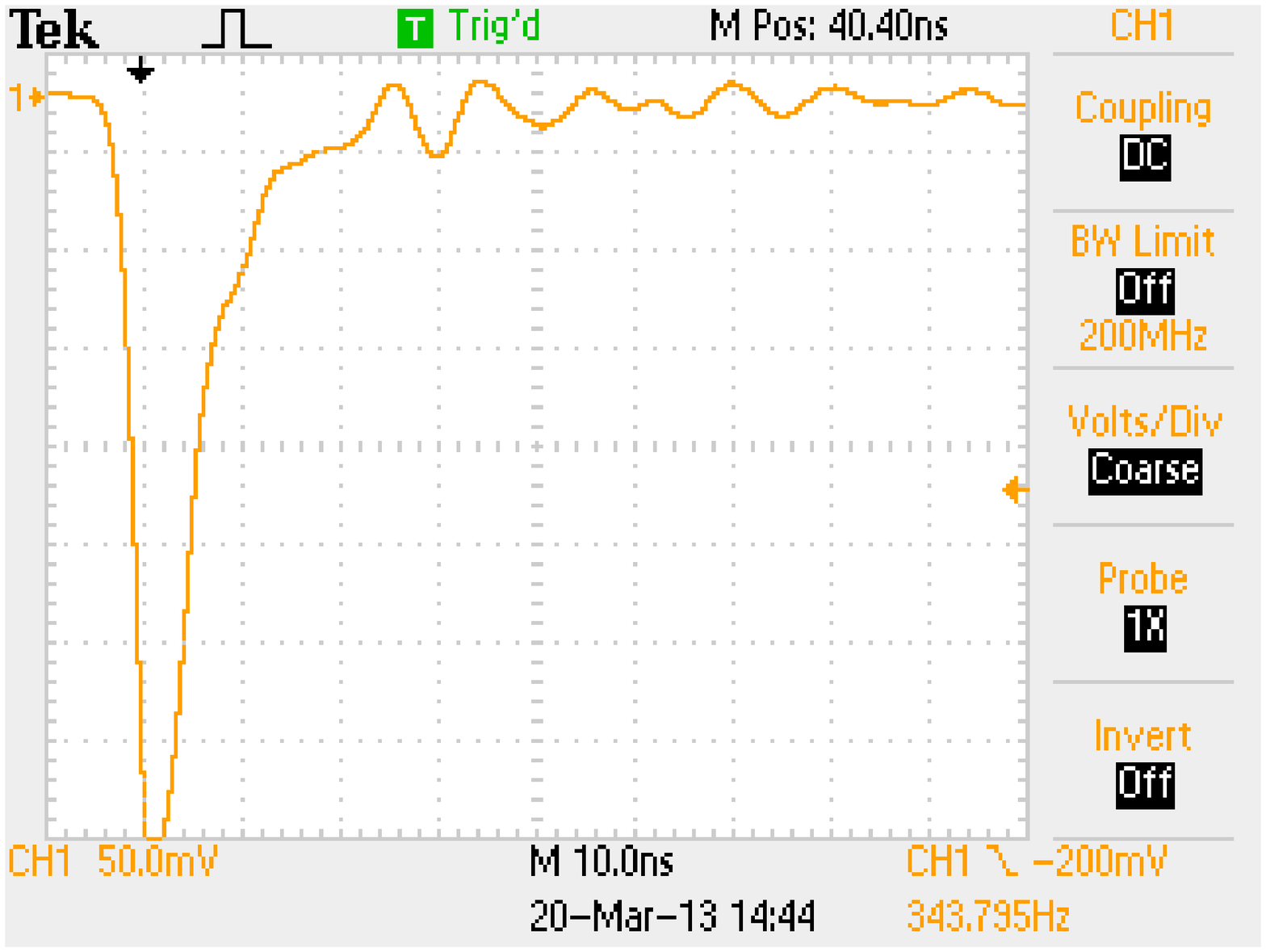}}
\caption{Oscilloscope screen shot of the H6528 PMT signal at 2000~V as received from the factory. (a) is an example of a discharge pulse, note that the vertical scale is 500~mV and time scale is 200~ns per division; (b) is an example of a normal pulse with ringing tails, note that the vertical scale is 50~mV and time scale is 10~ns per division. }
\end{figure}

Despite the high quality gains, the Hamamatsu PMT assemblies as received from the factory have some disturbing characteristics: all four detectors show frequent discharge pulses, see Fig~\ref{pic:discharge}; sizeable ringing tails are attached to primary pulse, see Fig~\ref{pic:ringing}. Regarding each of these characteristics, we developed customized methodology to overcome these issues. Hopefully, the solutions presented in this report will be useful to the manufacturer as well as other physicists who may be experiencing similar difficulties. 

\begin{figure}[t]
\centering
\includegraphics[width=0.70\textwidth]{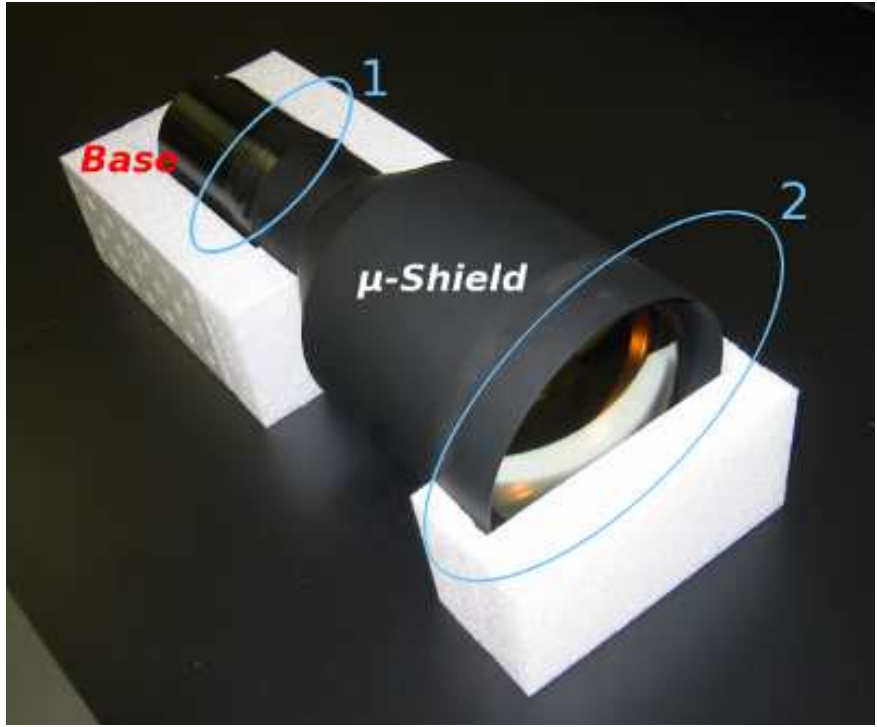}
\caption{H6528 PMT assembly as received from the manufacturer. Circles represent the identified discharge locations.}
\label{pic:pmt_factory}
\end{figure}

\section{Discharge Pulses}

After the H6528 was operated for 5-10~mins at a voltage higher than 1500~V, discharge pulses began to appear at a rate of few Hertz~(Hz) as observed on an oscilloscope. The discharge rate seems to be proportional to the operating voltage; the initial discharge rate at 2000 V is around 10~Hz, and gradually increases over time. Fig.~\ref{pic:discharge} shows an example of a discharge pulse. The height of the discharge pulse could vary from 200~mV to 10~V, and be a few $\mu$s wide.

Fig.~\ref{pic:pmt_factory} shows a H6528 as received from the manufacturer. Two locations were identified which have contributed the discharge pulses: 
\begin{enumerate}
\item 1st location is shown as the \#1 circle in Fig.~\ref{pic:pmt_factory}. Shield discharges to the base.
\item 2nd location is shown as the \#2 circle in Fig.~\ref{pic:pmt_factory}. Photo-cathode discharges to the shield. 
\end{enumerate}
From the our study, location \#1 contributes around 75\% of the discharge pulses.

\begin{figure}[t]
\centering
\includegraphics[width=0.7\textwidth]{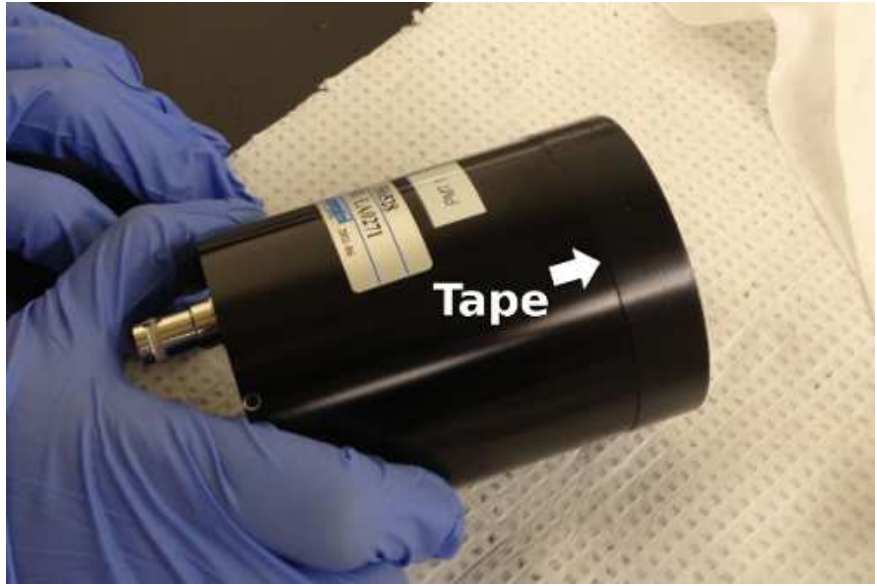}
\caption{H6528 base as received from the manufacturer. Note that one layer of electrical tape is at the end of the aluminum housing as white arrow indicates.}
\label{pic:base_orig}
\end{figure}

\subsection{Discharge from Shield to Base}

The basic design of the H6528 assembly consists of the PMT, the $\mu$-metal shield and the electronic base. The PMT tube slides into the metal shield from the larger opening of the shield; the aluminum shielded base is installed from the smaller opening of the shield. The pins at the end of the PMT need to be inserted into the socket on the base, which hold together the entire assembly. Note that a plastic retaining ring near the end of the smaller opening of the shield prevents light leak as well as centers the PMT inside the shield, as shown in Fig.\ref{pic:R1584}. 

\begin{figure}[t]
\centering
\includegraphics[width=0.62\textwidth]{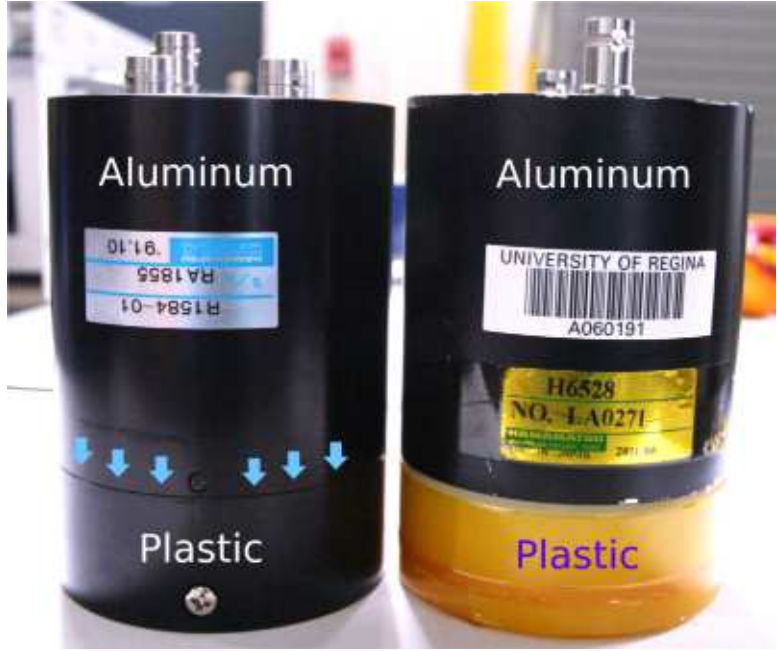}
\caption{Left: original R1584 PMT base was purchased by University of Virginia in 1991. Right: H6528 (current R1584) PMT base after modification; 1$^{\prime\prime}$ wide insulation ring is covered with 2 layers of kaptom tape.}
\label{pic:bases}
\end{figure}

In order to ensure the PMT pins are installed straightly into the socket, the base is designed to be inserted into the $\mu$-shield and pressed against the retaining ring, thus creating by 1~cm overlap between the base aluminum housing and $\mu$-metal shield. From the factory design, only one layer of electrical tape at the end of the base provides the electrical insulation between the base and $\mu$-mental shield, see Fig.~\ref{pic:R1584} and \ref{pic:base_orig}. Our study shows that this level of insulation is not sufficient to prevent electric discharge from the shield to the aluminum housing of the base. To completely insulate the shield from the base, the socket end of the aluminum housing of the base was trimmed off and replaced by a 1$^{\prime\prime}$ wide plastic ring, as shown in Fig.~\ref{pic:bases} (right).

Fig.~\ref{pic:bases} shows a modified H6528 base (right) and a R1584 base (left), which was purchased in 1991 by the University of Virginia. The photo was taken after all H6528 bases were modified. To our surprise, both bases both have a 1$^{\prime\prime}$ wide plastic ring to insulate the base from the shield. Obviously, Hamamatsu changed the 1991 base to the current design, thus causing discharges due to poor insulation.


\subsection{Discharge from Photo-Cathode to Shield}

The location of this discharge source is shown as the \#2 circle in Fig.~\ref{pic:pmt_factory}.  The discovery of this discharge issue was a surprise, since the $\mu$-metal shields are coated with insulating paint. Additional insulation on the inner surface of the shield is required to completely eliminate the discharge. However, the insulation material must be thin to allow PMT to slide through. The best option we found was to use 1-2 layers of 3/4$^{\prime\prime}$ kapton tape to cover 6~cm of the inner surface from the opening. Since the kapton tape has thickness of 25.4~$\mu$m  and a smooth surface, the PMT is able to slide freely into the shield.






\section{Ringing}

Fig.~\ref{pic:ringing} shows an example of the H6528 signal output as delivered from the factory. At 2000~V, the ratio between the primary and the first ringing tail is 10:1, this ratio increases as the operating voltage rises. When there is a large pulse ($>$1V), it is possible for the ringing tailing to trigger the electronic discrimination threshold, particularly at high event rates. It would be optimal to suppress the primary-to-ringing ratio without changing the other PMT signal output characteristics. After some study, we decided upon an electronic modification which satisfies the our requirement.

The modified H6528 base circuit diagram is attached in App. \ref{circuit}. Two 0.01~$\mu F$ high voltage capacitors are added between dynode chains \#10 and \#11 and the ground~\cite{semenov}. A 51 $\Omega$ metal film resistor is also introduced in series to the signal output; its power rating is 0.6~Watt.

\begin{figure}[t]
\centering
\includegraphics[scale=0.45, angle=90]{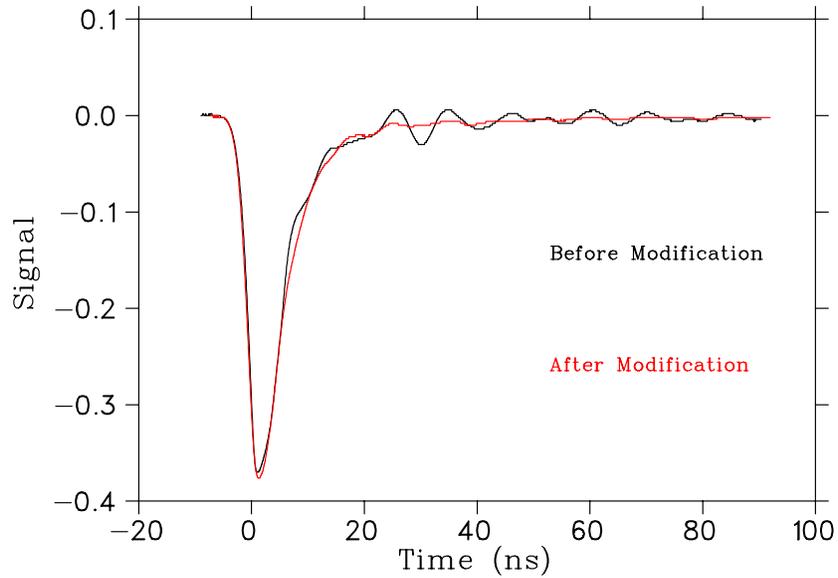}
\caption{PMT signal output before (black) and after (red) the modification at 2000~V. The oscilloscope trigger was set at 200~mV; both signals were averaged over 16 pulses.}
\label{pic:ringing_mod}
\end{figure}

Fig.~\ref{pic:ringing_mod} shows the original and modified PMT signal output at 2000~V. The ringing tails after the modification seem to be effectively suppressed to a satisfactory level.

\section{Test Results}

A series of tests were conducted to verify the performance of the H6528 PMT assemblies after modification. No discharges were observed on all four PMTs at a voltage up to 2500~V for a testing period of 3 days. Long term discharge monitoring will be implemented and documented. The PMT gains were repeated after the circuit modification and compared with the results from the initial measurement, the gains were found to be within 2\% of the original measurements, which is within our measurement error.

\section{Conclusion}

Despite the excellent gain performance, the Hamamatsu H6528 PMT assemblies have two disturbing characteristics: discharge pulses and ringing tails. In this report, we offer our solutions to overcome the issues: we introduced plastic insulation to stop discharge from the $\mu$-shield to the aluminum base; we used kapton tape to insulate the inner surface of the metal shield, thus stopping the discharge from the photo-cathode to the $\mu$-shield; we modified the base circuit to suppress the ringing tails effectively.

The modified PMT assemblies show no sign of discharge pulses so far, and the ringing tails are suppressed to a satisfactory level without significantly changing important characteristics of the PMTs.

\section{Acknowledgement}
We are grateful to A. Semenov, D. Gervais, D. Paudyal and D. Day for valuable suggestions and generous help to complete this work.

\pagebreak

\appendix

\pagebreak

\section{Circuit Modification for H6528 Tube}
\label{circuit}

The following is the modified circuit diagram of H6528 base.

\begin{figure}[h!]
\centering
\begin{pspicture}(2,0)(1.2,24.9)
\includegraphics[angle=90, scale=0.23]{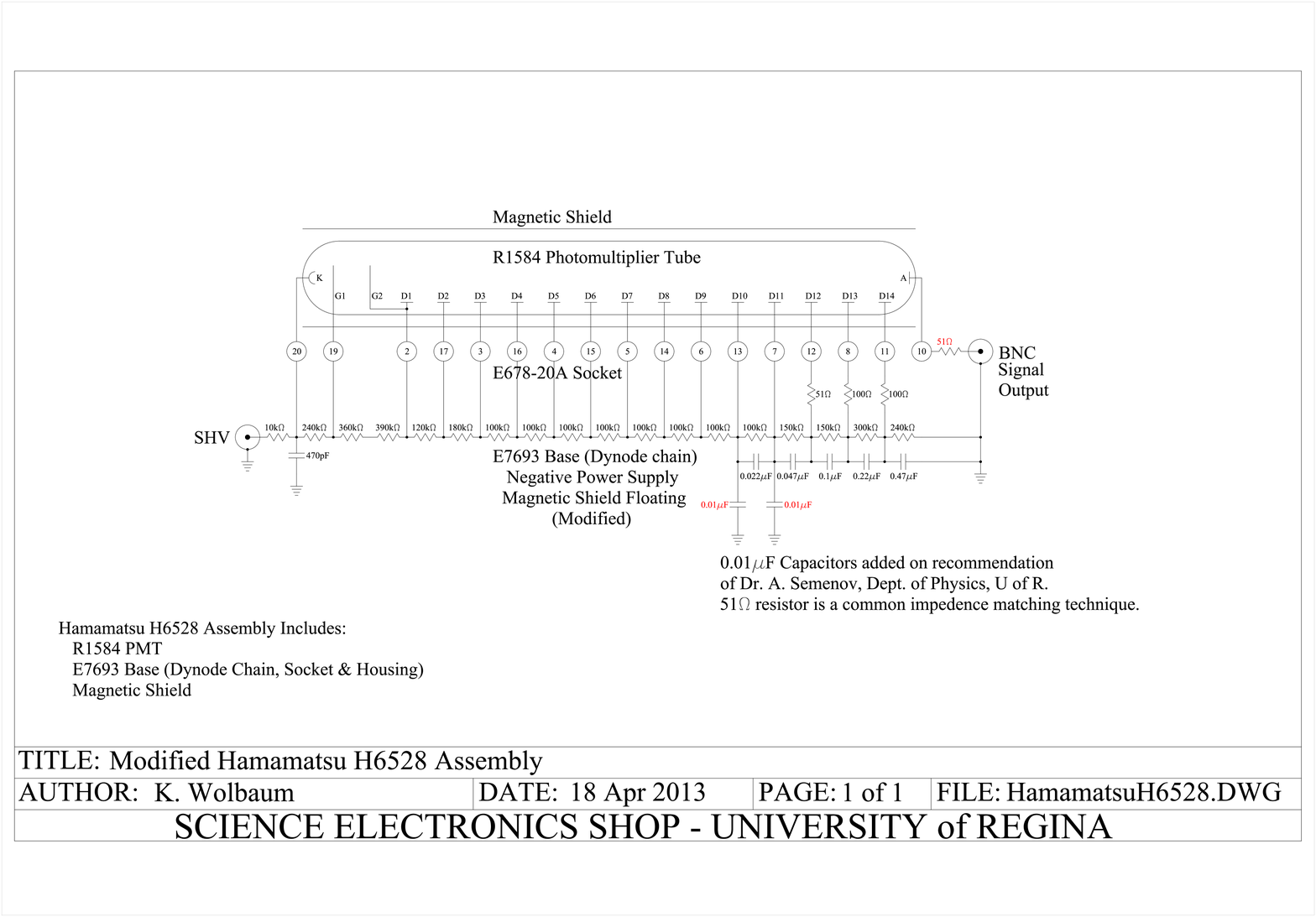}
\end{pspicture}
\end{figure}

\pagebreak


\begin{thebibliography}{00}

\bibitem{hamamatsu_cataloge} Hamamatsu Photomultiplier Tube Catalogue, www.hamamatsu.com/Photomultiplier.
\bibitem{alex} A. Fischer, G. M. Huber, \emph{Performance Testing of 5 inch PMTs}, HallC-doc-738-v1.
\bibitem{semenov} Private communication with Dr. A. Semenov.

\end{thebibliography}
\end{document}